\documentclass{article}

\usepackage{PRIMEarxiv}
\usepackage{rotating}
\usepackage[numbers,sort]{natbib}
\usepackage{url}
\usepackage{booktabs}
\usepackage{tikz-dependency}
\usepackage{pifont, natbib}

\usepackage[many]{tcolorbox}
\usepackage{tabularx}
\newtcolorbox{post_box}[3][]{%
boxsep=2.5pt,left=2pt,right=2pt,bottom=2pt,
width=\columnwidth,
boxrule=1pt,
colbacktitle=white,coltitle=black,
boxed title style={size=normal,colframe=white,boxrule=0pt}, 
interior style={white},
enhanced,
float,
fonttitle=\scshape,
title=Example~\thetcbcounter: #2,
#1
}
\usepackage{float}

\usepackage{tikz}
\usepackage{standalone}

\usepackage{pgfplots}
\pgfplotsset{compat=newest}
\usepgfplotslibrary{groupplots}
\usetikzlibrary{shadows,arrows.meta,positioning,backgrounds,fit,calc, fadings,shadows,shapes.arrows,intersections}

\tikzset{
    module/.style={
        draw, rounded corners,
        minimum width=80mm,
        minimum height=25mm,
        font={\fontsize{22pt}{20}\sffamily},
        text width=80mm, 
        align=center,
        anchor=center
        },
    line/.style={draw, color=black, -latex, thick, line width=2pt},
    back group/.style={
        rounded corners, draw=red, dashed, inner xsep=15pt, inner ysep=15pt
        },
    header/.style = {
        inner ysep = +2.2em,
        append after command = {
            \pgfextra{\let\TikZlastnode\tikzlastnode}
            node [header node] (header-\TikZlastnode) at (\TikZlastnode.north) {#1}
            }
        },
    header node/.style = {
        minimum width = #1,
        text depth    = +0pt,
        fill          = white,
        font={\fontsize{25pt}{20}\sffamily}
        },
    my arrow/.style = {
        single arrow, fill=red!50, 
        align=center, text width=1.5cm
    },
    my down arrow/.style = {
        single arrow, fill=red!50, 
        align=center, text width=1.5cm,
        rotate=-90
    },
    my left arrow/.style = {
    single arrow, fill=red!50, 
    align=center, text width=1.5cm,
    rotate=-180
    }
}
\pgfdeclarelayer{background}
\pgfdeclarelayer{foreground}
\pgfsetlayers{background,main,foreground}

\usepackage{amsthm}
\usepackage{amsmath,amsfonts}

\newcommand{\fqt}[1]{``#1''}
\newcommand{\fqtperiod}[1]{``#1.''\xspace}
\newcommand{\fqtcomma}[1]{``#1,''\xspace}

\newcommand{\fsl}{\textsl}

\DeclareMathAlphabet{\mathsl}{OT1}{ptm}{m}{sl}

\usepackage{diagbox}

\usepackage{xspace}

\newcommand{\citepos}[1]{\citeauthor{#1}'s \citeyearpar{#1}}

\newcommand{\ifsubmit}[1]{}

\newcommand{\be}{\begin{itemize}}
\newcommand{\ee}{\end{itemize}}
\newcommand{\bn}{\begin{enumerate}}
\newcommand{\en}{\end{enumerate}}
\newcommand{\bc}{\begin{center}}
\newcommand{\ec}{\end{center}}
\newcommand{\bl}{\begin{flushleft}}
\newcommand{\el}{\end{flushleft}}
\newcommand{\beq}{\begin{equation}}
\newcommand{\eeq}{\end{equation}}
\newcommand{\bq}{\begin{quote}}
\newcommand{\eq}{\end{quote}}

\newcommand{\bmp}{\begin{minipage}}
\newcommand{\emp}{\end{minipage}}

\DeclareMathAlphabet{\mathsl}{OT1}{ptm}{m}{sl}

\usepackage{xcolor}
\usepackage{siunitx}
\usepackage[np]{numprint}
\usepackage{multirow}
\npthousandsep{,}
\newcolumntype{T}{>{\tiny}l} 
\newcolumntype{H}{>{\Huge}l} 
\usepackage[inline]{enumitem}
\setlist[description]{leftmargin=1em}
\setlist[itemize]{leftmargin=1em}
\setlist[enumerate]{leftmargin=1.5em}

\usepackage{tikz}
\usepackage{standalone}
\usepackage{tikz-qtree}
\usepackage{makecell}

\usepackage{dcolumn}
\newcolumntype{d}[1]{D{.}{.}{#1}}
\makeatletter
\def\DC@endright{$\hfil\egroup\@dcolcolor\box\z@\box\tw@\dcolreset}
\def\dcolcolor#1{\gdef\@dcolcolor{\color{#1}}}
\def\dcolreset{\dcolcolor{blue}}
\dcolcolor{blue}
\makeatother
\makeatletter
\def\dcolreset{\dcolcolor{red}}
\dcolcolor{red}
\makeatother

\usepackage[normalem]{ulem}

\begin{document}
\title{The Blame Game:\\ Understanding Blame Assignment in Social Media}

\author{
    Ruijie Xi,
    Munindar P. Singh
    \\
    North Carolina State University \\
    Raleigh, North Carolina\\
    rxi@ncsu.edu, mpsingh@ncsu.edu
}


\maketitle

\begin{abstract}
Cognitive and psychological studies on morality have proposed underlying linguistic and semantic factors.
However, laboratory experiments in the philosophical literature often lack the nuances and complexity of real life.
This paper examines how well the findings of these cognitive studies generalize to a corpus of over 30,000 narratives of tense social situations submitted to a popular social media forum. 
These narratives describe interpersonal moral situations or misgivings; other users judge from the post whether the author (\textsl{protagonist}) or the opposing side (\textsl{antagonist}) is morally culpable.
Whereas previous work focuses on predicting the polarity of normative behaviors, we extend and apply natural language processing (NLP) techniques to understand the effects of descriptions of the people involved in these posts.
We conduct extensive experiments to investigate the effect sizes of features to understand how they affect the assignment of blame on social media.
Our findings show that aggregating psychology theories enables understanding real-life moral situations.
Moreover, our results suggest that there exist biases in blame assignment on social media, such as males are more likely to receive blame no matter whether they are protagonists or antagonists.

\end{abstract}


\section{Introduction}
\label{sec:introduction}
How do people judge whether someone deserves blame for their actions? 
This question has been extensively studied in social science. 
\citet{malle-2014-theory-blame} find that people assign blame to individuals they observe violating norms.
\citet{gray-2011-blame} show that victims can escape from being blamed, whereas heroes may cause blame.
\citet{guglielmo-2019-asymmetric} suggest that moral blame is more complex than moral praise.
Such social science experiments were conducted mainly through questionnaires and surveys, which stylize the social situations and limit the number of participants. 
In contrast, online social systems enable people worldwide to share viewpoints about a spectrum of social situations on various topics, allowing researchers to explore real-life blame with the aid of computational tools.

\begin{post_box}[label=first]{Sample post and comments on it.}

\textbf{Title:}\\
\fsl{``AITA for snitching on my sister?''}\medskip

\textbf{Body:}\\
\fsl{``\ldots I told my parents that my sister was staying up late with her tablet even though they had said she couldn't do it anymore. Now she's mad  \ldots  ''} \medskip

\textbf{Top-level Comment:}\\
\fsl{``\textcolor{red}{OTHER}. While you shouldn't be parenting her, you didn't go to your parents until she repeatedly ignored them as well as your warnings \ldots ''}
\medskip

\textbf{Top-level Comment:}\\
\fsl{``\textcolor{red}{AUTHOR}. If your sister was doing something really bad that hurt someone \ldots You have undermined her trust in you \ldots ''}
\medskip

\textbf{Flair: ``\textcolor{purple}{OTHER}''}
\end{post_box}

This paper examines a popular subreddit (i.e., forum), /r/AmITheAsshole (AITA),\footnote{\url{https://www.reddit.com/r/AmItheAsshole/}} where users describe interpersonal conflicts and other users (i.e., audience) comment and judge who deserves blame.
Example~\ref{first} shows a post and associated comments from AITA.
The \textsl{title} and \textsl{body} are of the post, \textsl{top-level comment} comes from the audience (often including a verdict), and \textsl{flair} is the verdict of the top-voted comment.
The most common verdicts are \textsl{author} and \textsl{other}.
We use the term \textsl{blame assignment} to represent a post's verdict.
Section~\ref{sec:dataset_coll} provides additional details.

Previous works take AITA as a resource for studying crowd-sourced blame assignments on first-person moral situations.
Much attention falls on accurately predicting verdicts \cite{lourie-2020-scruples, emelin-2020-moralstories, jiang-2021-delphi}.
These works apply neural networks such as transformers \cite{Devlin-2019-BERT} to obtain high accuracy of prediction performance. 
However, these models focus on accuracy but do not shed light on what linguistic characteristics affect the audience's decisions on assigning blame. 
Moreover, these models may be flawed since they don't consider social psychology research \cite{fraser-2022-does}. 


Previous empirical work has not studied how social psychology theories generalize to real-life situations posted in AITA.
According to the Theory of Dyadic Morality (TDM), blame is assigned to an \emph{agent} when behaviors are causing damage to a vulnerable \emph{patient} \cite{schein-2018-dyadic}.
Under TDM, agents are perceived as blameworthy, where their agentiveness depends on how they are described \cite{gray-2011-blame}.
The posts in AITA are first-person narratives that involve multiple individuals' and social identities (e.g., genders).
Accordingly, the audience assigns blame to the individuals that they think are described as agents. 
However, what descriptive features of individuals affect the audience's recognition of agentiveness remains unstudied.

This work studies the features of AITA's posts that affect the audience's blame assignment.
We especially focus on social psychology research relating to language and social features.
Language features affect social media data in many ways. For instance, \citet{beel-2022-linguistic} find sentiment is powerful in predicting the contentiousness of conversations on Reddit.
In addition, social factors, such as gender,  affect social media interactions \cite{beel-2022-linguistic} and can lead to biases.
For instance, \citet{candia-2022-demo} find that males have a higher possibility to receive blame on social media.
\citet{ferrer-2021-biasreddit} find that Reddit posts' topics are gender biased, for instance, \textsl{power}-related posts are associated with males.
Moreover, social scientists observe that gender and age affect morality in many ways \cite{wark-1996-gender, bracht-2018-moral}.
For instance, \citet{reynolds-2020-moraltypecasting} find that moral typecasting stereotypes females into the role of suffering patients.

\citet{malle-2014-theory-blame} proposes that assigning blame is a cognitive process that requires individuals to foresee the negative outcomes of agentive behaviors.
Therefore, we define \textsl{cognitive-affective features} as language features that can shape the audience's blame assignment decisions.
To this end, this paper aims to address two research questions:
\begin{description}
    \item[RQ\textsubscript{Feature}:] What cognitive-affective language features are crucial in blame assignment?
    \item[RQ\textsubscript{Social}:] What biases, if any, arise in blame assignment on social media?
\end{description} 

\subsection{Methods}
To answer RQ\textsubscript{Feature}, we operationalize a set of novel factors using Natural Language Processing (NLP) that have explanatory power.
We propose a novel \textbf{entity-centric} approach that partitions individuals involved in a situation as the protagonist (author) and antagonists (others).
Then, we collect language features describing the entities based on existing social science research, such as emotions conveyed from attributive and predicative words \cite{mohammad-2013-emotion}.
The language features are categorized into contextual, psycholinguistic, and linguistic features, where psycholinguistic features are entity-based and others are situation-based. 
Although previous research uses these features to analyze social media data \cite{hoover-2020-mft, sap-2017-connotation}, it doesn't apply them to morality with entity-based approaches.
We use the proposed features to build machine learning classifiers to predict whether an entity will receive blame.
Using these classifiers, we examine the features' effect sizes to understand how an entity causes blame given the description.

To answer RQ\textsubscript{Social}, we consider gender and age as social factors that may lead to biases in blame assignment \cite{botzer-2022-analysis, candia-2022-demo}.
We extend the previous works by conducting qualitative and quantitative analysis using a post's textual information, which helps understand a situation in linguistic terms.
We extract the demographics of the entities from the posts (they are marked with expressions such as [25F]).
We apply statistical methods to measure the association strengths between blame assignment and these social factors.

\subsection{Contributions and Findings}
This paper contributes in two aspects.
First, we characterize blame assignment with novel features inspired by social psychology literature.
We show these features have sufficient accuracy in predicting blame assignment while being interpretable.
Second, our proposed methods go beyond theoretical models and provide insights that can benefit psychological research, such as optimizing language used in surveys for laboratory experiments of studying morality.  

\begin{table*}[htb]
    \centering
    \caption{Summarizing recent work on moral-decision making models and AITA.}
    \begin{tabularx}{\textwidth}{p{3.2cm} p{2.7cm} p{5.6cm} X}
    \toprule
        Type&Paper &  Description &Dataset\\
        \midrule
        Moral-decision making&\citet{lourie-2020-scruples} & Building neural models to predict moral scenarios& Scraped from AITA\\
        &\citet{forbes-2020-social} & Building neural models to predict morality of social norms& Crowd-sourced dataset\\
        &\citet{emelin-2020-moralstories} & {Building neural models to predict intents, actions, and consequences of social norms}&Crowd-sourced dataset \\
        &\citet{jiang-2021-delphi}& Building neural models to provide moral advisor& Multi-sourced datasets\\
        \midrule
        Statistical Analysis&\citet{nguyen-2022-mapping}& Taxonomizing the structure of moral discussions& Scraped from AITA\\
        &\citet{candia-2022-demo}& Analyzing demographic information of blame assignments& Scraped from AITA\\
        &\citet{zhou-2021-assessing}& Analyzing linguistic features in blame assignments& Scraped from AITA\\
        &\citet{botzer-2022-analysis}& Analyzing morality by building a moral judgment classifier& Scraped from AITA\\
        \bottomrule 
    \end{tabularx}
    \label{tab:comparison}
\end{table*}
Our analyses show that certain generic linguistic characteristics are highly correlated with blame assignment across the board: for instance, the protagonist can reduce blame by eliciting positive perspectives (e.g., supportive) towards themselves.
Additionally, authors can reduce blame when they describe themselves using less powerful words, whereas using dominance-related words triggers blame.
Furthermore, our analysis suggests that authors in the 15--45 age group are more likely to attract bias than others.
In addition, males have a higher possibility to receive blame whether they are protagonists or antagonists, especially when they post specific situations, such as discussing \textsl{medicines and medical treatment} and \textsl{judgment of appearance}.

\subsection{Literature Review}
Recently, researchers have considered the
possibility of improving moral decision-making through AI by understanding (im)moral social norms and behavioral rules using NLP.
Table~\ref{tab:comparison} summarizes recent relevant research, in two groups.
The first group deals with predicting moral judgments. 
\citet{lourie-2020-scruples} use the title of a post in AITA as a social norm and develop a large dataset including human described situations based on the social norms. 
\citet{forbes-2020-social} break down blame assignments of one-liner scenarios into rules of thumb and ask annotators to write moral and immoral stories based on the rules.
Similarly, \citet{emelin-2020-moralstories} build crowd-sourced dataset based on the social norms \cite{lourie-2020-scruples}, which include actions, intentions, and consequences of a moral situation. Other works build moral judgment classifier to apply to other social media \cite{botzer-2022-analysis} and predict blame assignment on one-line natural language snippets from possibilities such as understandable, wrong, bad, and rude \cite{jiang-2021-delphi}.

However, social scientists point out that the general enterprise of training morality models on crowd-sourced data with no underlying moral framework is deeply flawed, as is the case for the Delphi system \cite{jiang-2021-delphi} \cite{talat-2021-word, fraser-2022-does}.
Besides, psychologists note the necessity for such AI systems to have a coherent understanding of human moral psychology \cite{liu-2022-advisor}.
Hence, we do \emph{not} aim to build the most accurate judgment predictor---and thus do not compete with state-of-art neural models.
Instead, we expand the computational modeling of moral understanding based on how social psychology constructs are apparent in language.

The second group in Table~\ref{tab:comparison} concerns analyzing AITA using statistical methods, such as creating a taxonomy of moral discussions \cite{nguyen-2022-mapping}, analyzing the correlation between users' demographics and blame assignments \cite{candia-2022-demo}, and identifying linguistic features in moral judgment \cite{zhou-2021-assessing}.
However, no work has studied the effects of the descriptions on individuals' agentiveness reflected in social media.

\section{Proposed Method}
\label{sec:method}
\begin{figure}[!ht]
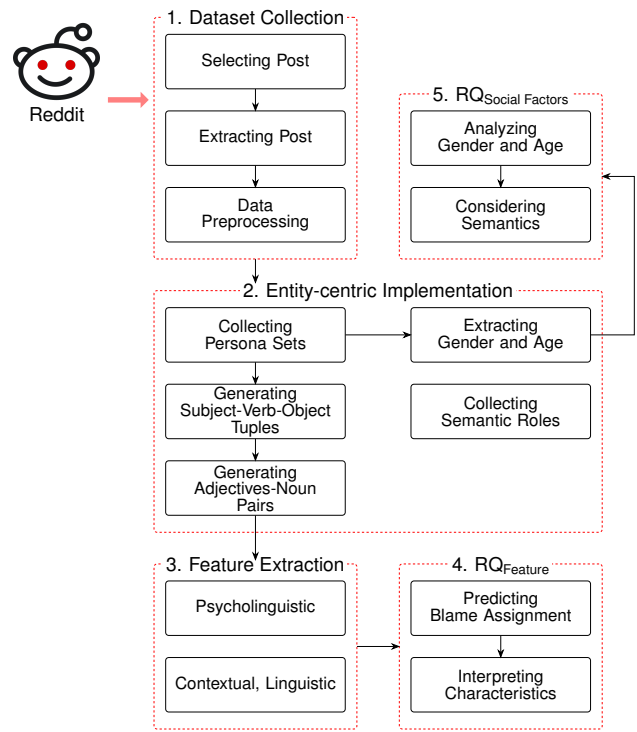

    \centering
    \includestandalone[width=0.5\columnwidth]{./tikz/framework}
    \caption{Overall pipeline of the proposed method.}
    \label{fig:framework}
\end{figure}

Our framework is divided into four phases as shown in Figure~\ref{fig:framework}. 
\begin{enumerate}
\item \textsl{Dataset collection} involves collecting data from a subreddit and preprocessing the data into a proper format.
\item \textsl{Entity-centric implementation} involves separating entities, generating subject-verb-object (SVO) tuples, collecting semantic roles, extracting gender and age, and generating adjectives-noun pairs (ANP).
\item \textsl{Feature Extraction} includes measuring psycholinguistic, contextual, and linguistic features.
\end{enumerate}

\subsection{Dataset Collection}
\label{sec:dataset_coll}
\subsubsection{Selecting posts from Reddit}
Although AITA has been used in previous studies, they are either not public \cite{zhou-2021-assessing, candia-2022-demo} or insufficient for answering our research questions \cite{lourie-2020-scruples, nguyen-2022-mapping}.
Our work needs the selected posts to contain predicate-argument structures that previous works haven't mentioned.
To improve relevance and accuracy, we constrain our dataset (FAITA) (details are in Section~\ref{sec:method}) to include posts that have:
\be{}
    \item Been given flairs (the determined verdict). 
    \item At least 50 top-level comments (judgments).
    \item Majority votes the same as the flair.
    \item At least ten extractable subject-verb-object tuples and ten extractable adjectives-noun pairs.
\ee{}

\subsubsection{Extracting Post}
We use the PushShift API\footnote{\url{https://github.com/pushshift/api}} and Reddit API\footnote{\url{https://www.reddit.com/dev/api}} to extract data over July 2020--July 2021.
Some AITA stories may be faked to solicit outrage.
The moderator deletes posts that are not truthful or not about interpersonal conflicts, which violate the subreddit rules.\footnote{\url{https://www.reddit.com/r/AmItheAsshole/about/rules/}}
We remove undesirable posts---those deleted, from a moderator, or too short---to ensure that the posts in our dataset decrease the conflicts between two parties and avoid discrepancies between data from Reddit and the archived data from PushShift.

Each judgment in the comments takes the form of a code: YTA, NTA, ESH, NAH, and INFO, which correspond to the classes AUTHOR, OTHER, EVERYONE, NO ONE, and MORE INFO. 
However, labeling the post with the majority votes may be inaccurate because morality is relative.
Instead, we extract the \textsl{title}, \textsl{text}, and \textsl{flair} of each post.
The \textsl{Flair} of each post is determined by the verdict of the top-voted comment {18} hours after submission (or the majority computed from its ten top-level comments if there is no flair field).
We assign labels to YTA as 1 and NTA as 0 and discard other codes.
This process yields 32,696 posts. We randomly split posts into 80\% as the training, 10\% as the development (dev), and 10\% as the test sets.
Table~\ref{tab:dataset} shows the distributions of FAITA.

\begin{table}[!htb]
    \centering
    \begin{tabular}{l r r r}
        \toprule
         Dataset&Train&Dev&Test\\
         \midrule
         {Posts}&26,156&3,270&3,270\\
         {Sentences}&376,846&125,766&125,332\\
         {Author Wrong (label 1)}&9,874&1,182&1,238\\
         {Others Wrong (label 0)}&16,282&2,088&2,031\\
        \bottomrule
    \end{tabular}

    \label{tab:dataset}
\end{table}

\subsubsection{Data Preprocessing}
We combine the title and text of posts in FAITA.
We preprocess the text using the NLTK toolbox.\footnote{\url{https://www.nltk.org/}}
We remove all emojis, punctuation (except periods for separating sentences), symbols, and special characters and replace contractions with patterns (e.g., replace \textsl{can't} with \textsl{can not}). 
We tokenize the sentences and lemmatize tokens using WordNet Lemmatiser \cite{poria-2012-wordlemma}.
We identify a \fqt{sentence} as \textsl{words separated by a period in the original post}.

\subsection{Entity-Centric Implementation}
We build a set of syntax-aware methods for extracting the protagonist (author) and antagonist (others) of each post using entity coreference and the syntactic dependency parse. These \textbf{entity-centric methods} require partitioning entity tokens into the protagonist and antagonist persona sets, understanding how the authors are portrayed the \fqt{cast of main characters} in the narratives, and how these characters behave.
We use Semantic Role Labeling (SRL) \cite{jurafsky-2009-nlp-book} to identify the protagonist and antagonist in each post.

\subsubsection{Collecting Persona Sets}
The protagonist and \textsl{antagonists} persona sets, respectively, contain first-person pronouns (e.g., I, me, and we), and third-person pronouns (e.g., she, he, and they).
We add the pronouns to the persona sets as key tokens.
We use the Spacy\footnote{\url{https://spacy.io}} dependency parser to extract more candidate terms by identifying part of speech tags (e.g., PRON, PROPN, and NOUN).
We filter the nouns and proper nouns by a total of 3,125 people-related words from prior research, such as characters in history textbooks \cite{li-2020-historytextbook-bias}.
Thus, we can collect all the people-related nouns and proper nouns.
Then we use Huggingface\footnote{\url{https://huggingface.co}} \texttt{neuralcoref} for coreference resolution, and append all tokens from spans that corefer to the pronouns in protagonist or antagonist persona sets, respectively.

\subsubsection{Collecting SRL}
Unlike syntax-aware methods, SRL analyzes sentences with respect to predicate-argument structures such as \fqtperiod{\textsl{who} did \textsl{what} to \textsl{whom} and \textsl{when} and \textsl{how} and \textsl{why}}
We employ the AllenNLP BERT-based Semantic Role Labeller \cite{gardner-2018-allennlp} to extract spans tagged \texttt{ARG0} for \textsl{agent} and \texttt{ARG1} for \textsl{patient}.
As the following example shows, each sentence may have multiple tagged spans; thus, we first identify the SRL-tagged sentences.
\bn{}
   \item \textbf{They} (\texttt{ARG0}) claimed \textbf{me} (\texttt{ARG1}) a dependent even though \textbf{I} (\texttt{ARG0}) have been financially independent for about a year.
\en{}
In each post, we match the entities in persona sets with SRL labels \texttt{ARG0} or \texttt{ARG1}.
This enables us to find when the author describes themselves or others as \textsl{agent} or \textsl{patient} in each narrative.

\subsubsection{Generating Subject-Verb-Object (SVO) Tuples}
\begin{figure}[!htb]
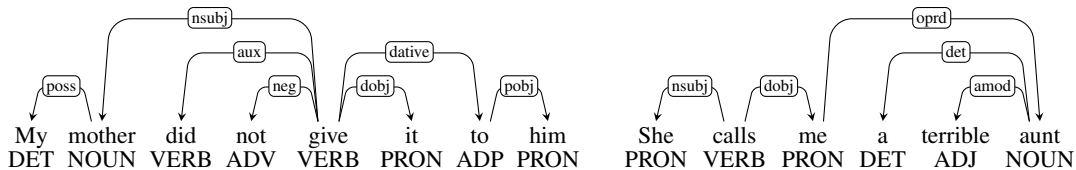

\small
    \centering
    \begin{dependency}
        \begin{deptext}[column sep=0.1em]
             My \& mother \& did \& not \&[0.3em] give \&[0.3em] it \& to \& him\\
              DET \& NOUN \& VERB \& ADV \& VERB \& PRON \&ADP \& PRON \& \\
        \end{deptext}
          \depedge{2}{1}{poss}   
          \depedge{5}{2}{nsubj}
          \depedge{5}{3}{aux}
          \depedge{5}{4}{neg}
          \depedge{5}{6}{dobj}
          \depedge{5}{7}{dative}
          \depedge{7}{8}{pobj}
    \end{dependency}
    \begin{dependency}
        \quad 
        \begin{deptext}[column sep=0.2em]
            She \& calls \& me \& a \& terrible \& aunt \& \\
            PRON \& VERB \& PRON \& DET \& ADJ \& NOUN \& \\
        \end{deptext}
          \depedge{2}{1}{nsubj}   
          \depedge{2}{3}{dobj}
          \depedge{3}{6}{oprd}
          \depedge{6}{4}{det}
          \depedge{6}{5}{amod}
    \end{dependency} 
    \caption{Dependency parsers for the example sentences.}
    \label{fig:tree}
\end{figure}

Beginning from the persona sets, we first identify verbs (VERB) that have dependencies with entities in the persona sets using a syntactic dependency parse tree.
We consider entities typed \texttt{nsub} (nominal subject), \texttt{nsubjpass} (passive nominal subject), \texttt{csubj} (clausal subject), \texttt{csubjpass} (passive clausal subject), \texttt{xsubj} (controlling subject), to the verbs as subjects; we consider entities typed \texttt{dobj} (direct object), and \texttt{iobj} (indirect object), to the verbs, as objects.
Then we add the SVO tuples for persona sets accordingly.
Besides the directly generated SVO tuples, we also generate new SVO tuples by finding entities from spans that corefer to the subject or object.
Using a dependency parser, it is possible to handle the negation of the verbs and add a \fqt{not} before the verb as shown in Figure~\ref{fig:tree}.

Figure~\ref{fig:tree} shows two sentences and their dependency trees in one post.
Here, we consider four people, \textsl{my mother, him, she, me}, and two SVO tuples, which are \textsl{(my mother, not give, it)} and \textsl{(she call me)}.
The coreference resolution finds \textsl{she} corefers to \textsl{my mother}, so we add \textsl{(my mother, call, me)} to the SVO tuples.

\subsubsection{Generating Adjectives-Noun Pairs (ANP)}
Adjective-noun pair is a semantic construct for capturing the effect of an adjectival modifier to modify the meaning of the nouns such as \fqt{cute dog} or \fqtperiod{beautiful landscape}
Similarly, we use a dependency parse tree to identify adjectives for the entities in the persona sets.
We use \texttt{amod} (adjectival modifier), \texttt{acomp} (adjectival complement), and \texttt{ccomp} (clausal complement) dependencies to select the adjectives modifying the entities.
As shown in Figure~\ref{fig:tree}, after we add \textsl{aunt} to the protagonist persona set, we find \textsl{terrible, aunt} pair because of the \texttt{amod} tag.
We also add \textsl{terrible, me} because \textsl{aunt} corefers to \textsl{me}. 

\subsubsection{Extracting Gender and Age for Persona Sets}
\label{sec:extract_persona}
Note that posts on AITA are interpersonal stories; the complexity of the description makes it difficult and expensive to extract the social factors of the antagonist.
Therefore, we extract and explore the social factors of the protagonist to improve the accuracy of the generated social factors.
Gender and age identities are not typically available on Reddit, allowing for anonymous posting.
Fortunately, the social media template for posting gender and age, e.g., [25f] (25-year-old female) or \texttt{(?i:i|i am a)([mf]|(?:fe)?male))}, enables us to use regular expressions to extract the information. 
We extract age by string matching on the gender modifier or the numeric age.
To improve the accuracy of the extraction, we consider two string matching patterns: \textsl{I [25m] and my wife [25f]}.
Besides, we consider gendered alternatives where available; for example, male can be estimated by \texttt{$\backslash$b(boy|father|son)$\backslash$b}) and female by (\texttt{$\backslash$b(girl|mother|daughter)$\backslash$b}).
We do not match nonbinary genders because we do not have ground truth labels for nonbinary targets.
To evaluate the regular expression, we took a random sample of 300 submissions and checked the result. 
We found no false positives and 2\% false negative cases. 
In addition, when gender and age are extracted by regular expression, it matches the manually labeled one 94\% of the time.

\subsection{Feature Extraction}
\label{sec:features}
We categorize the features into \textbf{contextual}, \textbf{psycholinguistic}, and \textbf{linguistic} features.
We measure the psycholinguistic features for subject-verb-object tuples and adjectives-noun pairs of the protagonist and antagonist persona sets separately.
We calculate scores for other features of each post.
Table~\ref{tab:features} summarizes the categories.

\begin{table*}[htb]
    \centering
    \caption{Feature categories and explanations.}
    \begin{tabularx}{\textwidth}{p{2cm} p{5.0cm} X}
    \toprule
        Category & Feature & Explanation \\
        \midrule
        Contextual & Topic & Lexicon-based topics measured by LDA \cite{blei-2003-lda}; each post has a list of topic it belongs to  \\
        Contextual & Content & TF-IDF weighted n-grams (n=1,2)\\
        \midrule
        Psycholinguistic & Agent versus Patient & Ratio of author and others being an \textsl{agent} or a \textsl{patient} \cite{schein-2018-dyadic}\\
        Psycholinguistic & Connotation Frames & Scores of connotation frames-related \cite{sap-2017-connotation} words normalized by count of subject-verb-object tuples; the scores are calculated separately as writers' perspective, value, effect, mental state \\
        Psycholinguistic & Agency and Power & Agency and power scores \cite{rashkin-2016-connotation} normalized by count of subject-verb-object tuples\\
        Psycholinguistic & Moral Content & Occurrences of the five virtue-vice paired Moral Foundation Theory lexicon \cite{Hopp2020TheEM} normalized by count of subject-verb-object tuples and count of adjectives-noun pairs\\
        Psycholinguistic & Valence, Arousal, Dominance (VAD)  & Occurrences of VAD lexicon \cite{mohammad-2018-vad} normalized by count of subject-verb-object tuples and count of adjectives-noun pairs\\
        Psycholinguistic & Emotion  & Occurrences of Emotion lexicon \cite{mohammad-2013-emotion} normalized by count of subject-verb-object tuples or count of adjectives-noun pairs.\\
        \midrule
        Linguistic & Subjectivity & Occurrences of subjectivity-related words \cite{wilson-2005-recognizing} normalized by count of words\\
        Linguistic & Hedge & Occurrences of hedge words \cite{hyland-2018-metadiscourse} normalized by count of words\\
        Linguistic & Modal & Occurrences of modal words normalized by count of words\\
        Linguistic & Pronoun & Occurrences of first, second, and third pronouns \\
        Linguistic & Sentiment & Averaged VADER \cite{hutto-2014-vader} compound scores; nominal sentiment categories\\
    \bottomrule
    \end{tabularx}
    \label{tab:features}
\end{table*}

\subsubsection{Contextual Features}
\label{sec:cont_feats}
Content is essential in analyzing social media posts \cite{guo-2019-cmv, zhou-2021-assessing}. 
We extract the content at two levels: term frequency-inverse document frequency (TF-IDF) weighted n-grams vectors (n = 1, 2) and post-level topics. 
TF-IDF weights combine term frequency $tf(t, d)$ (the occurrence of a term $t$ in a document $d$) and inverse document frequency $idf(t, D)$ (the rating of $t$ in a corpus $D$).
It reflects how important a word is to a document in a corpus.

\begin{table}[!htb]
    \centering
    \caption{Sample topics with representative words.}
    \begin{tabular}{p{0.30\columnwidth} p{0.60\columnwidth}}
        \toprule
        Topic Label & Top Weighted Words \\
        \midrule
        Relationship with family (20.8\%) & life, relationship, mother, ex, child, father, life, wife, partner, son\\
        \midrule
        Intimate relationship (17.3\%) & girlfriend, boyfriend, relationship, dating, upset, feel, pretty, lot, love, guy \\
        \midrule
        Living in shared accommodation (16.5\%)&apartment, rent, live, room, living, house, lease, stay, bedroom\\
        \midrule
        Money (7.3\%)&pay, rent, saving, buy, job, account, car, loan, afford, cost\\
        \midrule
        Pregnancy concerns in pets (5.5\%) & dog, child, husband, child, pregnant, puppy, cat, law, animal, birth\\
        \midrule
        Work (4.4\%) & hour, work, boss, company, manager, job, employee, office, shift, week\\
        \midrule
        Appearance judgment (4.2\%) & hair, look, wear, white, black, comment, clothes, dress, looked, pretty\\
        \midrule
        Neighborhood (3.3\%) & neighbor, phone, email, post, account, people, use, street, yard, facebook\\
        \bottomrule
    \end{tabular}
    \label{tab:topics}
\end{table}

We extracted topics using Latent Dirichlet Allocation (LDA) \cite{blei-2003-lda}.
LDA is a generative statistical model that assumes that each document (here, post) contains a distribution of topics; each topic is a distribution of words.
We train a model on the \textsl{text} of posts in FAITA, exploring the number of topics ranging over 30--55 and finalized on 30, as it achieves the lowest perplexity.
We then combine the topics that contain fewer than 200 posts.
Table~\ref{tab:topics} shows a sample of eight hand-selected topics and ten example words belonging to each topic.
In Table~\ref{tab:topics}, the \fqt{Topic Label} column is summarized manually by the authors according to all the words they are associated with; the percentage indicates how frequently each topic occurs in the dataset. 
We also show the ten most representative words for each topic.
These topics show that posts in FAITA range from family to work issues.
Additional topics with at least 100 posts include: driving safety (2.8\%), gender communication differences (2.8\%), games (2.6\%), cooking (2.7\%), holiday gifts (2.7\%), social media (2.3\%), wedding plan (2.1\%), medical treatment (1.6\%), and school (1.1\%).


\subsubsection{Psycholinguistic Features}
These refer to the lexico-semantic analysis of the cognitive association that a word carries and its literal meaning.
The scores being introduced are separated into \textsl{agent}, \textsl{connotation frames}, \textsl{power and agency}, \textsl{moral content}, and \textsl{VAD}.
From SVO tuples, we calculate scores for entities as subjects and objects.
From ANP, we calculate scores for entities based on their adjective modifier.
We normalize the scores calculated for entities in the persona sets to capture the values of the protagonist and antagonists.

\paragraph{Agent} The ratio of the time the protagonist and antagonists are assigned as \textsl{agents} or \textsl{patients} in a post using SRL labels.

\paragraph{Connotation Frames} A formalism for analyzing subjective roles and relationships implied by a given predicate \cite{rashkin-2016-connotation}.
To analyze nuanced dimensions of narratives in FAITA, we draw from a lexicon with annotations for 1,000 most frequently used English verbs across various dimensions, ranging from {--1} to {1}.
A verb might elicit a positive sentiment for its subject but imply a negative sentiment for its object.
For example, from \fqtcomma{\textsl{Alice} betrayed \textsl{Bob}} the annotation contains the following dimensions:

\be{}
    \item Writer's perspective. The writer (protagonist) elicits a negative perspective toward \textsl{Alice} as {--0.67} (e.g., blaming) and a positive perspective toward \textsl{Bob} as {0.26} (e.g., supportive).
    \item Reader's perspective. (1) Values: the reader presupposes a positive value of \textsl{Bob} as {0.87} (strongly positive) and \textsl{Alice} as {0.47} (neutral to positive). (2) Effects: the reader presupposes the harms towards \textsl{Bob} as {--0.93} (strongly negative) compared to \textsl{Alice} as {0.067} (neutral). (3) Mental states: the reader presupposes \textsl{Bob} is most likely to feel negative ({--0.67}) as a result of the event, but \textsl{Alice} it not likely to be affected (--0.03).
\ee{}

\paragraph{Power and Agency} A pragmatic formalism organized using frame semantic representations \cite{sap-2017-connotation} to model how different levels of power and agency are implicitly projected on people through their actions. 
We use \citepos{sap-2017-connotation} extension lexicon of Connotation Frames to measure the agency and power scores of \textsl{author} and \textsl{others}.
This extension lexicon contains more than 2,000 transitive and intransitive verbs to model how different levels of power and agency are implicitly projected on the entities through their behaviors. 
Entities with high agency (subjects of \textsl{attack}) are active decision-makers, whereas entities with low agency (subjects of \textsl{doubts} and \textsl{needs}) are passive.
This lexicon contains binary labels of each verb, which are positive ({1}), equal ({0}), and negative ({--1}). 

\paragraph{Moral Content}  
The Moral Foundation Theory \cite{haidt-2007-mft} has been widely adopted in the computational social community, which is critical in understanding how the psychological influence of social content unfolds, such as quantifying moral behaviors in Twitter \cite{hoover-2020-mft} and taxonomizing the structure of moral discussions in Reddit \cite{nguyen-2022-mapping}. 
We adopt the extended Moral Foundations Dictionary (eMFD) \cite{Hopp2020TheEM}, which is a crowdsourced dictionary-based tool for extracting moral content from textual corpora.
The eMFD contains 2,041 unique words, which are categorized into five broad domains based on MFT: care/harm, fairness/cheating, loyalty/betrayal, authority/subversion, and sanctity/degradation.
Each word in the dictionary has a composite valence score ranging from {--1} to {1}.

\paragraph{VAD (Valence, Arousal, Dominance)} 
The three affective dimensions are used to measure affective meanings from words that convey the author's attitudes toward the events and people referenced.
We obtain the valence scores for 20,000 words from the NRC VAD lexicon \cite{mohammad-2018-vad}, which contains real-valued scores ranging from {0} to {1} for each category.


\paragraph{Emotions} 
Emotions conveyed in words represent sentiment from the authors toward the described entities \cite{mohammad-2013-emotion}, which may place a considerable cognitive load on the audience \cite{dijkstra-1995-character}.
The NRC Emotion lexicon \cite{mohammad-2013-emotion} provides the emotion of around 20,000 words, indicating whether a word is associated with an emotion category.
The categories are joy, sadness, anger, fear, trust, disgust, surprise, and anticipation.

\subsubsection{Linguistic Features}
We estimate linguistic scores for \textsl{subjectivity}, \textsl{hedge}, \textsl{sentiment}, and \textsl{modal} in each post.
\paragraph{Subjectivity} arises when people express personal feelings or beliefs, e.g., in opinions or allegations \cite{wilson-2005-recognizing}, which comprises the authors' perspectives towards the descriptive situations, contributing to the audience's judgments.
We compute the subjectivity of a post as the average score of words based on the Subjectivity lexicon \cite{wilson-2005-recognizing} (nonneutral words of \fqt{weaksubj} = 0.5 and \fqt{strongsubj} = 1).
Additionally, we count the numbers of first-person, second-person, and third-person pronouns because words such as \fqt{you} and \fqt{we} engage the audience with the discourse.

\paragraph{Hedge} is associated with indirection in politeness theory \cite{brennan-1999-hedge}, which may affect the audience's judgments.

\paragraph{Sentiment} indicates emotions by conveying the polarity of an opinion. 
A negative tone may imply more immorality than a neutral tone.
We calculate each post's compound sentiment scores and sentiment categories with the VADER package \cite{hutto-2014-vader}.

\paragraph{Modal} words affect the sentiment of the words they modify \cite{kiritchenko-2016-sentiment}.

\section{RQ{\textsubscript{Feature}}{}: Blame Assignment Analysis}
\label{sec:result_analysis}
\noindent 
To answer RQ{\textsubscript{Feature}}, we perform two statistical analyses: (1) prediction---can description frames predict blame assignment? (2) language characteristics analysis---can linguistic features affect blame assignment? 
Here, we focus on using the prediction as a tool for analyzing, not for the purpose of making an accurate prediction.
Note that we do not take gender and age as features when conducting experiments as they are not available in some of the posts in FAITA.

\subsection{Predicting Blame Assignment}
\label{sec:verdict_prediction}
Now we examine how well computational models
can predict blame assignments in moral situations.
For machine learning models, we explore two logistic regression models (LR) to compute the probability of
a positive label for each sentence.
All the models are built using scikit-learn\footnote{\url{https://scikit-learn.org/stable/modules/generated/sklearn.linear_model.LinearRegression.html}} toolkit in Python.
An LR classifier computes the probability of a discrete outcome given an input variable.
BERT-LR is logistic regression where our features are replaced with the BERT \cite{Devlin-2019-BERT} embeddings of input sentences.
We evaluate the performance of different models in terms of recall, precision, and F1 scores.
All computation models were run 10 times and we measure the standard deviation of the scores for each method.
For LR, we set the class weights to \fqt{balanced} to account for the label imbalance.
And we explore feature selection using the L1-norm and regularization using L2-norm.
Other hyperparameters for LR include setting the weight ranging over (\SI{1e-4}, \SI{1e-3}, \SI{1e-2}, \SI{1e-1}).
We propose two baseline models.
Random predicts the verdict randomly. 
Length predicts using the lengths of the sentences in a post, which has been shown to be effective in predicting blame \cite{zhou-2021-assessing}.

The quantitative results of our methods are shown in Table~\ref{tab:result1}.
BERT-LR and LR outperform the baselines significantly, while BERT-LR performs
best. 
\begin{table}[htb]
    \centering
    \caption{Prediction accuracy (macro-average scores). All scores have standard deviations between 0.01 and 0.03. The best scores are in bold. We only report LR and BERT-LR results because they yield the performances of other models such as multilayer perceptron, SVM, and random forest.}
    \begin{tabularx}{\columnwidth}{ l X X X X X X }
        \toprule
            Method&
            \multicolumn{2}{c}{F1}&
            \multicolumn{2}{c}{Recall}&
            \multicolumn{2}{c}{Precision}\\
            &DEV&TEST&DEV&TEST&DEV&TEST \\
           \midrule 
            Random&0.49&0.50&0.50&0.50&0.50&0.49 \\
            Length&0.39&0.38&0.50&0.50&0.49&0.52\\ 
            \midrule            LR&0.66&0.65&0.65&0.64&\textbf{0.66}&\textbf{0.65} \\
            (\ding{53}) Linguistic&0.63&0.62&0.60&0.61&0.63&0.62 \\
            (\ding{53}) Contextual&0.53&0.53&0.54&0.54&0.60&0.60 \\
            (\ding{53}) Psycholinguistic &0.48&0.49&0.52&0.53&0.59&0.59 \\
            BERT&\textbf{0.71}&\textbf{0.72}&\textbf{0.68}&\textbf{0.69}&\textbf{0.66}&\textbf{0.65} \\
        \bottomrule
    \end{tabularx}
    \label{tab:result1} 
\end{table}
It is worth noting that our features, despite the lower performance than BERT-LR, are clearly informative of morality prediction because they directly capture the information contributing to the audience's decision on blame.
Transformer models such as BERT encode linguistic characteristics in a more sophisticated manner and may include additional information.
But it is less clear exactly what transformers capture and whether they capture irrelevant statistics.
To examine the contribution of each feature category, we conducted ablation tests based on the LR model. 
Regarding F1 scores, psycholinguistic features have the highest contribution, followed by contextual and linguistic. 
This result reaffirms the importance of analyzing the lexical semantics of attributive and predicative words in first-person moral narratives.

\subsection{Interpreting Characteristics}
We measure the effect size and statistical significance of each feature.
The effect of each feature is conditioned on the domain of each post using logistic regression. 
For interpretation purposes, we use the Odds Ratio (OR) (the exponent of the effect size).
Odds represent the ratio of the probability of an author being blamed to the probability of not being blamed; OR is the ratio of odds when the effect size increases by one unit.
The OR is calculated using the equation $OR = exp(\beta_i), i \in N$, where $\beta_i $ is the coefficient of attribute $i$ obtained by the trained LR model and $N$ denotes the attribute set.
Moreover, we estimate statistical significance by Spearman's Rank Correlation Coefficient to avoid assuming normality or other distributions for FAITA.

\subsubsection{Contextual Features}
We begin by looking at OR between topics and blame assignments.
Table ~\ref{tab:attr_topics} shows the OR values and correlation coefficients for the authors being blamed corresponding to Table~\ref{tab:topics}.
These results show that posts related to relationships, pregnancy concerns in pets, and games are positively correlated with blame assignment. 
Other topics mentioned in Section~\ref{sec:cont_feats} that may decrease the probability of an author being blamed are school, holiday gifts, and cooking; the rest are positively correlated.

\begin{table}[!htb]
\centering
\caption{Odds ratio (OR) and Spearman's correlation coefficient of topics calculated from the test set. 
An effect is positive (blue) if OR $>$ 1 and negative (red) if OR $<$ 1.}
    \begin{tabular}{l 
                S[table-format=-1.2] 
                S[table-format=-1.2] }
        \toprule
        Topic & {Moral Blame} & {P-value}\\
        \toprule
        Relationship with family&
        \color{blue}1.11&0.02\\
        Intimate relationship &\color{blue}1.07&0.07\\
        Living in shared accommodation&\color{red}0.79&0.02\\
        Money &\color{red}0.82&0.20\\
        Pregnancy concerns in pets &\color{blue}1.46 & 0.03\\
        Work &\color{red}0.98&0.03\\
        Appearance judgment&\color{blue}1.16& 0.07\\
        Neighborhood  &\color{red}0.71 & 0.14\\
        \bottomrule
        
    \end{tabular}
    \label{tab:attr_topics}
\end{table}

\subsubsection{Psycholinguistic Features}
Table~\ref{tab:attr_psycho-lin} shows the features that influence at least a 1\% probability of the author being blamed.
Table~\ref{tab:attr_psycho-lin} reveals that psycholinguistic features are informative.
The results for agent and patient are consistent with social psychology that \fqt{being a victim can help escape blame} \cite{gray-2011-blame}.
These features do not store lexical information but affect the audience's judgments in the cognitive aspect.
In addition, our analysis indicates the protagonist can reduce blame by eliciting positive perspectives (e.g., supportive) towards themselves. 
Additionally, authors can reduce blame when they describe themselves as suffering more from harm than the antagonist.
The Agency and Power features are consistent with the above findings because the high values imply the agent's high-level authority and powerful capability, which can trigger blame.

Care and harm are opposite concepts in Moral Foundation Theory, whereas increasing the use of words from the lexicon reduces the probability of the author being blamed.  
Moreover, we find that VAD features do not have significant p-values.
However, they increase the probability of the author being blamed by 23\% when increasing the use of the dominance lexicon when describing the protagonist.  
Different emotion categories have different effects on blame assignment.
Specifically, using sadness-related words when describing the protagonist lowers the probability of the authors being blamed to one-third.
Additionally, increasing the use of disgust-related words when describing the antagonist more than doubles the probability of the author being blamed.
We highlight a possible explanation: the description frames of the protagonist and antagonist need to be captured as a whole, not as individual components.

\begin{table}[!htb]
\centering
    \caption{Odds ratio (OR) and Spearman's correlation coefficient of psycholinguistic features calculated from the test set. WP represents \textsl{writer's perspective}.
    An effect is positive (blue) if OR $>$ 1 and negative (red) if OR $<$ 1.}
    \centering
    \begin{tabular}{l S[table-format=-1.2] 
                S[table-format=-1.2] S[table-format=-1.2] 
                S[table-format=-1.2] }
        \toprule
        Feature&\multicolumn{2}{c}{Protagonist}&\multicolumn{2}{c}{Antagonist}\\
        \cmidrule(lr){2-3}
        \cmidrule(lr){4-5}
        {}&{Moral Blame OR}&{p-value}&{Moral Blame OR}&{p-value}\\
        \midrule
        Agent& \textcolor{blue}{1.93} &0.05 & \textcolor{red}{0.93}&0.002\\
        Patient& \textcolor{red}{0.53} &0.03 & \textcolor{blue}{1.01}&0.001\\
        \midrule
        WP& \textcolor{blue}{1.01} &0.006&\textcolor{red}{0.81}&0.031\\
        Value& \textcolor{red}{0.99} &0.13 & \textcolor{blue}{1.02}&0.13\\ 
        \midrule
        Power & \textcolor{blue}{1.04} &0.006&\textcolor{red}{0.97}&0.003\\
        Agency & \textcolor{blue}{2.00} &0.003 &\textcolor{red}{0.96}&0.002\\
        \midrule
        Care&\textcolor{red}{0.99}&0.03&\textcolor{red}{1.03}&0.03\\
        Harm& \textcolor{red}{0.97} &0.08& \textcolor{blue}{1.00}&0.07\\
        Betrayal & \textcolor{red}{0.95} &0.06&\textcolor{red}{1.11}&0.16\\
        Loyalty &\textcolor{red}{0.97}&0.08&\textcolor{blue}{1.03}&0.17\\
        \midrule
        Valence& \textcolor{red}{0.99} &0.14&\textcolor{blue}{1.22}&0.13\\
        Arousal& \textcolor{blue}{1.04} &0.11&\textcolor{blue}{1.21}&0.15\\
        Dominance& \textcolor{blue}{1.23} &0.14 &\textcolor{blue}{1.09} &0.15\\
        \midrule
        Joy &\textcolor{red}{0.98} &0.09&\textcolor{red}{0.98}&0.13\\
        Sadness&\textcolor{red}{0.31}&0.05&\textcolor{blue}{1.28}&0.005\\
        Anger &\textcolor{blue}{1.05} &0.01&\textcolor{red}{0.11}&0.03\\
        Fear &\textcolor{blue}{2.33}&0.01&\textcolor{blue}{1.06}&0.03\\
        Trust&\textcolor{blue}{1.10}&0.08&\textcolor{red}{0.20}&0.09\\
        Disgust&\textcolor{blue}{1.06}&0.07&\textcolor{blue}{2.16}&0.02\\
        Anticipation &\textcolor{blue}{1.34} &0.05&\textcolor{blue}{1.74}&0.04\\
        \bottomrule
    \end{tabular}
    \label{tab:attr_psycho-lin}
\end{table}

\subsubsection{Linguistic Features}
As shown in Table~\ref{tab:attr_lin}, subjectivity is positively correlated to blame assignment in contrast to hedging, which indicates that subjective descriptions increases the possibility of the author being blamed with greater certainty.
The frequent use of third-person pronouns triggers blame because the audience may think that the author is trying to escape from blame by avoiding describing themselves.
Although second-person pronouns have a small p-value, they increase the probability only by 1\% of the author being blamed. 
However, the negative sentiment category strongly affects blame assignment with an OR of 3.18, which may explain that extreme sentiment triggers blame assignment.

\begin{table}[!htb]
\centering
    \caption{Odds ratio (OR) and Spearman's correlation coefficient of linguistic features calculated from the test set. 
    An effect is positive (blue) if OR $>$ 1 and negative (red) if OR $<$ 1.}
    \begin{tabular}{l S[table-format=1.2] 
                S[table-format=1.2]}
        \toprule
        Feature & {Moral Blame} & {p-value}\\
        \midrule
        Subjectivity & \textcolor{blue}{1.09} & 0.006\\
        Hedge& \textcolor{red}{0.66} &0.04\\
        First pronoun & \textcolor{blue}{1.45} &0.10\\
        Second pronoun& \textcolor{blue}{1.01} &0.0009\\
        Third pronoun& \textcolor{blue}{1.96} &0.003\\
        Sentiment score &\textcolor{blue}{1.78}&0.001\\
        Sentiment: positive &\textcolor{blue}{1.18} &0.005\\
        Sentiment: neutral &\textcolor{red}{0.99}&0.08\\
        Sentiment: negative &\textcolor{blue}{3.18} &0.01\\
        \bottomrule
    \end{tabular}
    \label{tab:attr_lin}
\end{table}

\section{RQ{\textsubscript{Social:}} Social Factors Analysis}
\label{sec:analysis_gender_bias}
In this section, we examine gender and age features in FAITA to investigate whether audiences exhibit differences in their assessments of moral situations.

\subsection{Analyzing Gender and Age Association}
\label{sec:analysis_gender_bias_1}
This section investigates whether the authors' self-reported gender and age lead to an imbalance in blame assignment. 
Using the method of Section~\ref{sec:features}, 13,935 posts describe the genders of all entities involved, and 6,079 posts state the authors' age is between 15 and 65.
To determine the association between blame and social factors, we perform the $\chi^{2}$ significance test and compute Cramer's $\phi$ as the effect size. Here, {0.07}--{0.21}, {0.21}--{0.35}, and $>${0.35} respectively indicate small, moderate, and strong association \cite{cohen1988_no_edition}.

We aggregate occurrences of entities being blamed when they are protagonists and antagonists.
The overall $\chi^{2}$ test result between genders and blame assignment is ($\chi^{2}(13,935)=515.02, p < 0.001$) with $\phi = 0.17$.
Whereas the effect size indicates a small association between gender and blame assignment in FAITA, the evidence indicates there is an association between the two ($p < 0.001$). 
Besides, we observe that males are 53\% (the log-odds-ratio of occurrences when authors of different genders receive blame) more likely to receive blame.
The results allow us to discern the direction of the biases due to gender: male authors are more likely to be considered agentive no matter their position.
The observation coheres with previous psychological research that some sets of biases stereotype females into the role of suffering \textsl{patient} on social media \cite{reynolds-2020-moraltypecasting}.

To further investigate the correlation between blame assignment and age, we divide authors' ages (antagonists' ages are scarce) into four groups ranging from {15} to {55} as the range accounts for almost 80\% of active Reddit users.
Table~\ref{tab:chi_gender_age} illustrates blame assignment is associated with protagonists' ages when they are in the 15--45 age group ($p<0.05$), especially when authors are in the 36--45 age group ($\phi=0.18$).

\begin{table}[!htb]
    \centering
    \aboverulesep=0ex
    \belowrulesep=0ex
    \caption{The columns are age ranges. $N$ represents the number of corresponding posts.  $p <$ 0.05 indicates the age group and blame assignment are associated.(** : $p <$ 0.05, ***: $p <$ 0.001.)}
      \begin{tabular}{l r r r r}
        \toprule
        & \multicolumn{4}{c}{Age Ranges}\\
        \cmidrule{2-5}
        Metrics&15-25 & 26-35 & 36-45&46-55\\
        \midrule
        $N$&3,554&1,951&410&136\\
        $\chi^{2}$&76.56 (***) &50.89 (***)&13.46 (**)&2.96 ()\\
        {$\phi$}&0.15&0.16&0.18&0.15\\
        \bottomrule
    \end{tabular}
    \label{tab:chi_gender_age}
\end{table}

\subsection{Considering Semantics with Social Factors}
\label{sec:seman_gender}
We now examine how blame assignment differs between female and male protagonists in similar situations.
We employ pretrained sentence-BERT models \cite{reimers-2019-sentence-bert} to cluster the {13,935} posts based on semantic similarity.
We learn embeddings of the posts' \textsl{titles} as they serve as summaries of posts.
To remove the effect of gender-related tokens, we replace gender-identified words with \fqt{someone} using the resources mentioned in Section~\ref{sec:extract_persona}.

We adopt Hierarchical Density-Based Spatial Clustering of Applications with Noise (HDBSCAN) \cite{mcinnes-2017-hdensity} because no external references identify topic numbers in FAITA. 
Then we perform dimension reduction with Uniform Manifold Approximation and Projection for Dimension Reduction (UMAP) \cite{mcinnes-2018-UMAP} to alleviate the problem of sparse embeddings. 
We fine-tune parameters for HDBSCAN and UMAP models \cite{angelov-2020-top2vec} by increasing the model's Density-Based Clustering Validation (DBCV) score \cite{moulavi-2014-dbcv}. 
To enhance the semantic similarity in each cluster, we exclude the clusters containing fewer than {50} posts.
This process yields {7,248} posts that clustered into {47} groups, where the counts of posts in a cluster range from {51} to {712}.
We measure $\chi^{2}$ and $\phi$ to select the clusters where gender is strongly associated with blame assignment ($p$ $<$ {0.001} and $\phi$ $>$ {0.35}) and find six clusters include {1,162} posts.

To categorize the semantics of the clusters, we use the UCREL Semantic Analysis System (USAS) \cite{usas}, a framework for automatic semantic analysis and tagging of text, which is based on McArthur's Longman Lexicon of Contemporary English \cite{mcarthur-1982-longman}.
USAS has a multitier structure with 21 major discourse fields subdivided in fine-grained categories such as \textsl{People}, \textsl{Relationships}, and \textsl{Power}.
Using USAS, we label each cluster with the most frequent tag (or tags) among the highest TF-IDF-scored nouns from the posts.
We notice that the topics of the most gender-associated situations in FAITA corroborate previous work on categorizing language biases in Reddit \cite{ferrer-2021-biasreddit}.
For example, the most frequent gender-polarized situations on FAITA (ordered by frequency) are \textsl{kin}, \textsl{relationship: intimate/sexual}, \textsl{groups and affiliation}, \textsl{anatomy and physiology}, \textsl{work and employment}, \textsl{sports}, \textsl{games}, \textsl{money}, \textsl{medicines and medical treatment}, and \textsl{judgment of appearance}.
These tags account for the most discussed topics, as Table~\ref{tab:topics} shows.
It is important to note that our analysis of social factors in blame assignment is more suggestive than conclusive.
Our analysis suggests that social biases exist in social media posts, which influences blame assignment, at least in some topics.

\section{Discussions}
\label{sec:discussions}
This paper contributes to studying the nature of morality by assessing social psychology insights on descriptive real-life situations.
We incorporate a novel set of language features with machine learning models for predicting who is considered blameworthy. 
Statistical methods help visualize the effects of the features.
The effective prediction performance confirms the linkage between blame assignments and psychological observations on social media.
For example, entities described using less \textsl{care}-related words (a category from Moral Foundation Theory \cite{haidt-2007-mft}) are more likely to receive blame.
Furthermore, our findings suggest that gender and age are associated with blame assignment.
For example, males are more likely to receive blame than females; biases in blame assignments are more likely to be presented when protagonists are in the 15--45 age group.

Our results can be explained by the fact that people perceiving themselves as deserving blame are subject to feelings of guilt \cite{scott1971internalization}.
In our case, these feelings conveyed from social media posts may affect the audience's decision making on who is blameworthy.
In addition, psychological literature observes that social media might have typecasting towards male and female individuals \cite{ferrer-2021-biasreddit, reynolds-2020-moraltypecasting}.
However, people of different genders might be subject to different social pressures and thus be different in choosing the language to describe conflict \cite{asher2017gender}.
Our results are coherent with these observations, which reaffirms the significance of considering social psychology instruments in using computational methods to understand the nature of morality \cite{fraser-2022-does}. 

\subsection{Implications}
Our work contributes a new framework to demonstrate the significance of psychological theories in real-life situations, with theoretical and empirical guidelines to assist in studying morality.
First, our research provides novel language features based on social psychology that enable textual and psychological insights into the nature of morality. 
Second, our proposed features can be applied to build interpretable models for blame assignment.
Practically, this work could motivate the design of future AI systems to incorporate psychological findings to promote the interpretability of real-world practical morality.

In addition, our study contributes to theoretical research such as The Theory of Dyadic Morality (TDM) \cite{schein-2018-dyadic} by providing language features.
For example, our analysis answers how individuals' agentiveness is affected by the associated descriptions.
Our work can help design laboratory experiments.
For example, since we demonstrate that language used to describe social situations may affect a participant's cognition, special attention could be paid when stylizing such situations.
Particularly, these designs can be tailored to subject-matter experts for studying advanced components in theoretical social research that demand human validations.

\subsection{Limitations and Future Work}
As in any study dealing with social media data, there are some limitations. 
First, this study design has the advantage of a higher ecologic validity but presents critical causal inference challenges.  
There might be hidden confounders that we cannot measure given the lack of data, such as the demographics of the audience.
In addition, the research may only be generalizable to some populations, which may not account for other factors influencing blame assignment on social media, such as age, culture, and education.
Therefore, the coefficients we find cannot be interpreted as a direct causal effect, which means our research is more suggestive than conclusive.

This work could be extended in interesting ways.
One direction is to incorporate language features from the comments accompanying each post to extract cognitive-affective features directly from the audience.
The accompanying comments may help explain \textsl{what}, \textsl{why}, and \textsl{how} language features affect the audience's cognitive processes.
In addition, to provide a causal explanation of how social factors appear in blame assignment and how they function, future work can leverage explicit and implicit social factors in the narratives.

\bibliographystyle{IEEEtranSN}
\bibliography{Ruijie}

\end{document}